\documentclass[a4paper]{jpconf}
\usepackage{graphicx}
\usepackage{siunitx}

\usepackage{lineno}

\begin{document}
\title{Measurement of ambient neutrons in an underground laboratory at Kamioka Observatory and future plan}


\author{Keita~Mizukoshi$^1$, Ryosuke~Taishaku$^1$, Keishi~Hosokawa$^2$, Kazuyoshi~Kobayashi$^{3, 4}$, Kentaro~Miuchi$^1$, Tatsuhiro~Naka$^{5, 6}$, Atsushi~Takeda$^{3, 4}$, Masashi~Tanaka$^7$, Yoshiki~Wada$^8$, Kohei~Yorita$^9$, Sei~Yoshida$^{10, 11}$ }

\address{$^1$ Department of Physics, Kobe University, Kobe, Hyogo 657-8501, Japan}
\address{$^2$ Research Center for Neutrino Science, Tohoku University, Sendai, Miyagi 980-8578, Japan}
\address{$^3$ Kamioka Observatory, Institute for Cosmic Ray Research, the University of Tokyo, Higashi-Mozumi, Kamioka, Hida, Gifu, 506-1205, Japan}
\address{$^4$ Kavli Institute for the Physics and Mathematics of the Universe (WPI), the University of Tokyo, Kashiwa, Chiba, 277-8582, Japan}
\address{$^5$ Nagoya University, Nagoya, Aichi 464-8602, Japan}
\address{$^6$ Kobayshi-Maskawa Institute for the Origin of Particles and the Universe, Nagoya, Aichi 464-8602, Japan}
\address{$^7$ Global Center for Science and Engineering (GCSE), Faculty of Science and Engineering, Waseda University, Shinjuku, Tokyo 169-8555, Japan}
\address{$^8$ Department of Physics, Faculty of Science, Tohoku University, Sendai, Miyagi 980-8578, Japan}
\address{$^9$ Department of Physics, Waseda University, Shinjuku, Tokyo 169-8555, Japan}
\address{$^{10}$ Department of Physics, Graduate School of Science, Osaka University, Toyonaka, Osaka 567-0043, Japan}
\address{$^{11}$ Project Research Center for Fundamental Sciences (PRC), Osaka University, Toyonaka, Osaka 560-0043, Japan}

\ead{mzks@stu.kobe-u.ac.jp}

\begin{abstract}
Ambient neutrons are one of the most serious backgrounds for underground experiments in search of rare events.
The ambient neutron flux in an underground laboratory of Kamioka Observatory was measured using a $\mathrm{^3He}$  proportional counter with various moderator setups.
Since the detector response largely depends on the spectral shape, the energy spectra of the neutrons transported from the rock to the laboratory were estimated by Monte-Carlo simulations. 
The ratio of the thermal neutron flux to the total neutron flux was found to depend on the thermalizing efficiency of the rock.
Thus, the ratio of the count rate without a moderator to that with a moderator was used to determine this parameter.
Consequently, the most-likely neutron spectrum predicted by the simulations for the parameters determined by the experimental results was obtained. 
The result suggests an interesting spectral shape, which has not been indicated in previous studies.
The total ambient neutron flux is $(23.5 \pm 0.7 \ \mathrm{_{stat.}} ^{+1.9}_{-2.1} \ \mathrm{_{sys.}}) \times 10^{-6}$ cm$^{-2}$ s$^{-1}$ \cite{ptep123}.
In this paper, we explain our method of the result and discuss our future plan.
\end{abstract}

\section{Introduction}\label{intro}

Ambient neutrons are one of the most serious backgrounds for future underground experiments, such as neutrinoless double beta decay searches, neutrino measurements, and direct dark matter searches.
The flux and the energy spectra are important information of underground laboratory.


We evaluated an ambient neutron spectrum with $\mathrm{^3}$He proportional counter at Kamioka Observatory \cite{ptep123}.
In the study, we considered the natural sources of ambient neutrons in a wall rock using data-driven Monte-Carlo (MC) simulation to estimate the shape of the neutron energy spectrum.
Considered natural sources were the ($\alpha$, n) reactions of $\mathrm{^{238}U}$ and $\mathrm{^{232}Th}$ series, spontaneous $\mathrm{^{238}U}$ fission, and cosmic muons interacting in the rock.
The generated neutrons were transported to the laboratory.
Consequently, the most likely energy spectrum was obtained.
In this paper, we will review the current method and propose a future plan to measure the comprehensive energy spectrum of ambient neutron in an underground laboratory.




%

\section{Current method}
\subsection{Detector setup} 
The study was performed with a $^3$He proportional counter which is widely used.
The counter (Model P4-1618-203 made by Reuter-Stokes Co.) has $\mathrm{^{3}He}$ gas at 10 atm and was made of a stainless steel cylinder (class SUS304), 38 cm in length and 5.18 cm in diameter.
The measurements were made in Lab-B at NEWAGE \cite{newage} experimental site, one of the underground laboratories at the Kamioka Observatory.
$\mathrm{^3He}$ has a large cross section to thermal neutrons (e.g., 5333 barns at 0.025 eV \cite{crosssection}), however, it is not sensitive to high-energy neutrons.
Moreover, information about the original kinetic energy of an incident neutron was lost because the Q-value of the reaction (0.764 MeV), was much larger than that of the detected neutrons.

To measure high-energy neutrons (in the MeV range), moderators and a shielding material were used.
A polyethylene moderator (outer radius of 9.9 cm, length of 51 cm, and thickness of 6.5 cm) was used to thermalize the high-energy neutrons so that they can be detected by the $\mathrm{^3 He}$ proportional counter.
An additional shielding material, a 4-mm-thick boron-loaded sheet (B~sheet), covered the moderator to reduce the effects of ambient thermal neutrons.
The B~sheet, of density 1.42 $\mathrm{g/cm^3}$, included 20 wt-\% $\mathrm{B_4 C}$, which shielded about 99.8\% of the thermal neutrons.

\begin{figure}[]
    \centering
    \includegraphics[width=10cm]{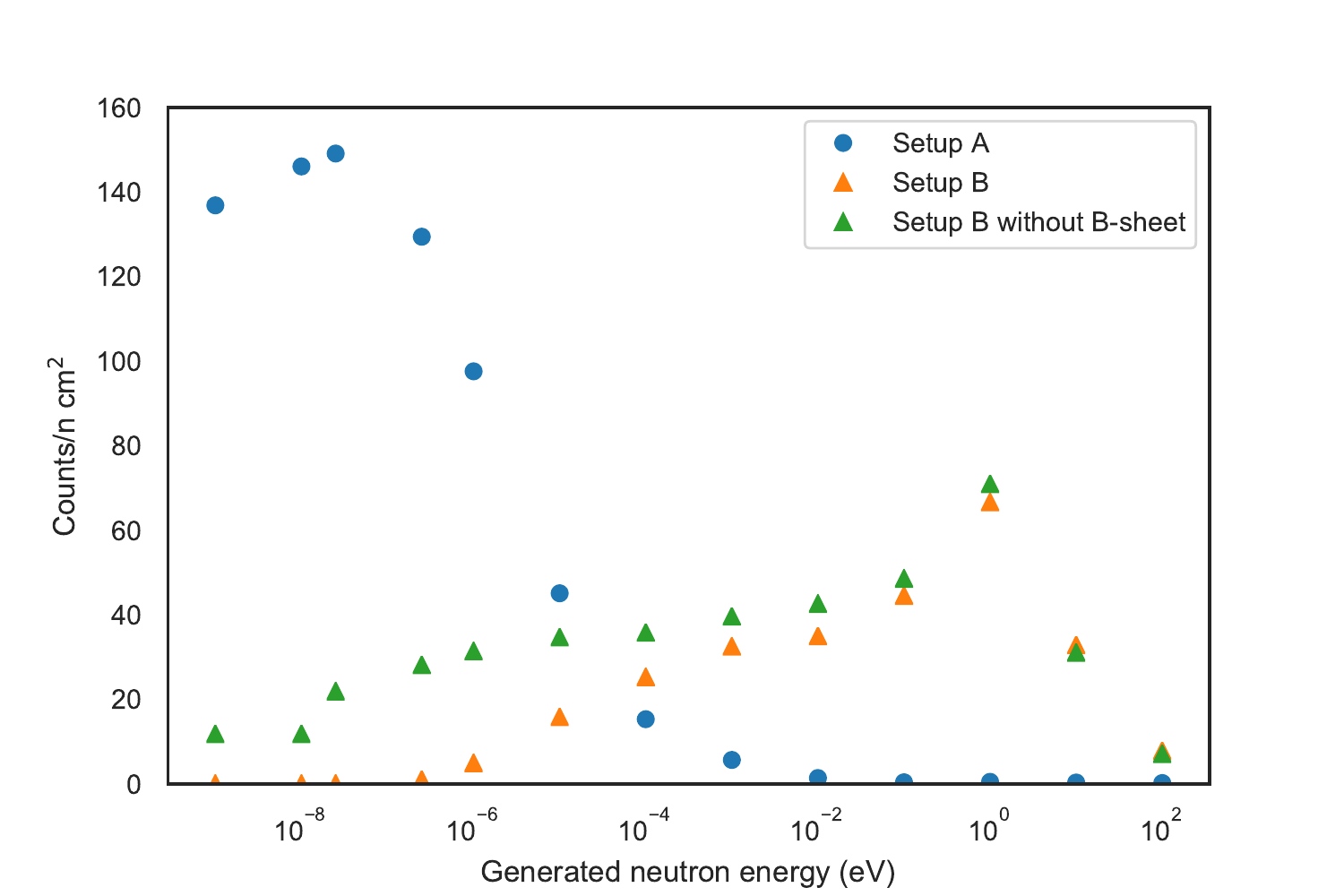}
    \caption{Expected counts of each setup for various neutron energies with a fluence of 1~neutron/cm$^2$ by Monte-Carlo simulation. The results of setup B without boron-containing sheet (B sheet) are also shown for reference.}
    \label{fig:eff}
\end{figure}

For setup~A, there was no moderator and no B~sheet.
Setup B had a moderator and the B sheet.
The detection efficiencies in each setup were evaluated by Geant4 \cite{geant4_1,geant4_2,geant4_3}.
Figure \ref{fig:eff} shows the expected numbers of counts when the neutron were generated with a fluence of 1 neutrons/$\mathrm{cm^2}$.
Setups A and B are mainly sensitive to thermal and fast (--MeV) neutrons, respectively.
The simulated result for setup~B without B sheet is also shown to illustrate the effect of the sheet.
It was observed that the B sheet shielded the ambient thermal neutrons, thereby making setup~B to be sensitive mainly to high-energy neutrons.
The simulation was checked by a $^{252}$Cf source calibration.



\section{Obtained results}\label{result1}

\subsection{Experimental results}
Table \ref{tab:result} summarizes the measurements obtained from the setups A and B.
The errors of a simple waveform cut, evaluation of total counts, and 5\% detector gain fluctuations including the electronics gain difference between setups were taken into account as systematic errors.

\begin{table}[]
\caption{Count rate and live time in each setup.}
\label{tab:result}
\centering
\begin{tabular}{lccccc}
    Setup &Start &Stop & Live time (day) & Rate ($\mathrm{10^{-3}}$cps)$\pm$$_\mathrm{stat.}$$\pm$$_\mathrm{sys.}$\\
\hline
    A  &Feb. 19 2016 &Mar. 20 2016  & 14.03 & 1.295 $\pm$ 0.034 $^{+0.039}_{-0.033}$\\
    B  &Oct. 19 2017 & Nov. 8 2017  &19.27  & 0.446 $\pm$ 0.018 $^{+0.013}_{-0.011}$\\
    \hline
\end{tabular}
\end{table}

\subsection{Simulation and calculation of total flux}
%

\begin{figure}[]
    \centering
    \includegraphics[width=10cm]{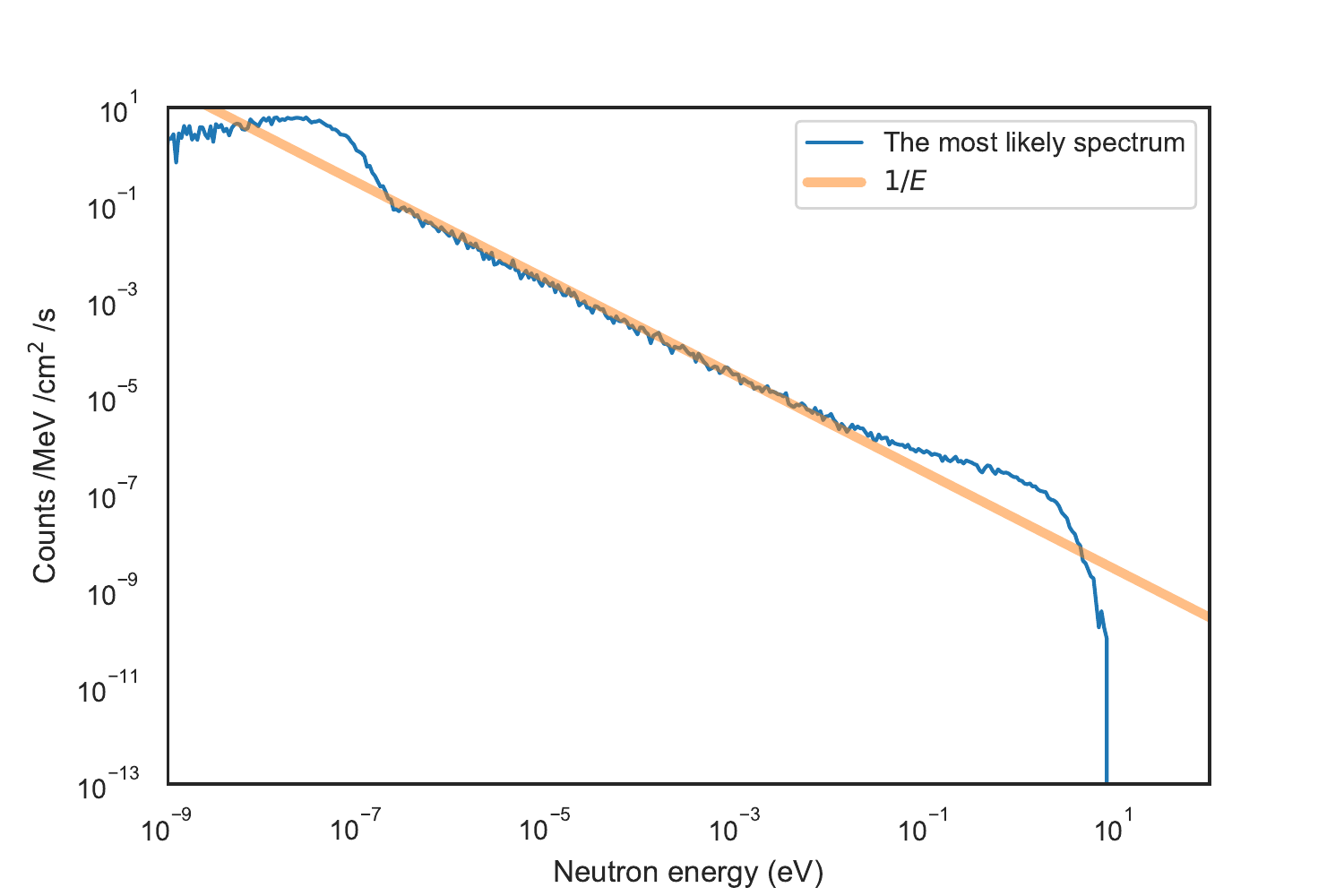}
    \caption{The most likely spectrum of ambient neutrons produced by ($\alpha$, n) reactions and spontaneous fission in Lab-B at Kamioka Observatory.}
    \label{fig:bestfit}
\end{figure}

To obtain the ambient neutron flux from the count rates, the shape of the energy spectrum was needed to be evaluated.
However, it is extremely difficult to derive the spectrum using only the $^{3}$He proportional counter alone.
Thus, we estimated the shape of the spectrum using MC simulation the considering origins of ambient neutrons and their transportation.
The dominant sources were ($\alpha$, n) reaction and spontaneous fission by Uranium and Thorium series nuclei in the rock of an experimental site.
Moreover, uncertain parameters for thermalization (rock components and MC thermalization) were determined so that to explain a ratio of the experimental rates (rate of Setup A / rate of Setup B).
We obtained the ambient neutron spectrum (Figure \ref{fig:bestfit}) using data-driven analysis.
The obtained spectrum was scaled by rate of Setup A.
The ambient neutron flux was $(23.5 \pm 0.7 \ \mathrm{_{stat.}} ^{+1.9}_{-2.1} \ \mathrm{_{sys.}}) \times 10^{-6}$ cm$^{-2}$ s$^{-1}$ with the spectrum.

\section{Future plan}

We obtained the energy spectrum and flux of an ambient neutron in an underground laboratory.
Most of the previous studies assumed that the spectral shape was a Boltzmann distribution and flat ($1/E$) for thermal and fast neutrons, respectively.
The results obtained in this research basically support these assumptions but also suggest an excess in the region from 100 keV to a few MeV.
The $^3 \mathrm{He}$ proportional counter is not sensitive in this region. 
Thus, other detectors are needed to increase our understanding of the spectrum structure.
We have started testing a low-background liquid scintillator to measure the high-energy neutron in a low-rate environment.

We are also interested in the position dependence of ambient neutron in the underground laboratory.
We are collecting data from multiple sites in the Kamioka Observatory using the $^3$He proportional counter, and the analysis is in progress.
As a future plan, we will observe long-term ambient neutron rate at a site to obtain a sign of annual modulation if it is.

\section{Conclusions}\label{conclusion}

Ambient neutrons are one of the most serious backgrounds for low-background experiments performed in underground laboratories.
However, neutron flux and energy spectrum are required to estimate the background precisely, so that it can be subtracted effectively.
The ambient neutrons were measured at Kamioka Observatory.
A $\mathrm{^3He}$ proportional counter was used to mainly detect thermal neutrons.
Higher energy neutrons were measured in different setups with different combinations of a polyethylene moderator and a B sheet.
The most likely energy spectrum was obtained by a data-driven MC simulation considering the origin of the neutron.
Using the spectrum, the total neutron flux was calculated to be $(23.5 \pm 0.7 \ \mathrm{_{stat.}} \ ^{+1.9}_{-2.1}\ \mathrm{_{sys.}}) \times 10^{-6} \ \mathrm{cm^{-2} s^{-1}}$.
These experimental results and our simulations suggest that there is an excess above $1/E$ in the MeV region in the ambient neutron spectrum.
The region should be measured directly by a detector which is sensitive to the energy range.
Meanwhile, we are preparing a low-background liquid scintillator.
We will continue our work to increase our detailed understanding of the ambient neutron at the underground laboratory.

\section*{Acknowledgments}
The authors thank Dr. Yuji Kishimoto of KEK for lending us the $\mathrm{^3He}$ proportional counter.
We also thank Kamioka Mining and Smelting Co., Ltd. for various supports given to our research activities in the underground laboratories.
 This work was supported by the MEXT KAKENHI Grant-in-Aid for Scientific Research on Innovative Areas 26104001, 26104003, 26104004, 26104005, JSPS KAKENHI Grant-in-Aid for Scientific Research (S) 24224007, JSPS KAKENHI Grant-in-Aid for Scientific Research (A)16H02189, (A)17H01661, and Grant-in-Aid for JSPS Fellows 19J20418.
 This work was partially supported by the joint research program of the Institute for Cosmic Ray Research (ICRR), the University of Tokyo. 
%


\end{document}